\documentclass[letterpaper]{article} %
\usepackage{aaai25}  %
\usepackage{times}  %
\usepackage{helvet}  %
\usepackage{courier}  %
\usepackage[hyphens]{url}  %
\usepackage{graphicx} %
\usepackage{natbib}  %
\usepackage{caption} %
\usepackage{algorithm}
\usepackage{algorithmic}

\usepackage{newfloat}
\usepackage{listings}
\usepackage{booktabs}
\DeclareCaptionStyle{ruled}{labelfont=normalfont,labelsep=colon,strut=off} %
\floatstyle{ruled}
\newfloat{listing}{tb}{lst}{}
\floatname{listing}{Listing}
\title{The Illusion of Empathy: How AI Chatbots Shape Conversation Perception}

\author{
    Tingting Liu\textsuperscript{\rm 1}\thanks{Corresponding author.},  
    Salvatore Giorgi\textsuperscript{\rm 1},  
    Ankit Aich\textsuperscript{\rm 1, 2}, 
    Allison Lahnala\textsuperscript{\rm 1}, \\
    Brenda Curtis\textsuperscript{\rm 1}, 
    Lyle Ungar\textsuperscript{\rm 2}, 
    João Sedoc\textsuperscript{\rm 3}
}
\affiliations{
    \textsuperscript{\rm 1}National Institute on Drug Abuse\\
    \textsuperscript{\rm 2}University of Pennsylvania\\
    \textsuperscript{\rm 3}New York University\\
\{tingting.liu, sal.giorgi, ankit.aich, allison.lahnala, brenda.curtis\}@nih.gov, ungar@cis.upenn.edu, jsedoc@stern.nyu.edu
}

\begin{document}

\maketitle

\begin{abstract}
As AI chatbots increasingly incorporate empathy, understanding user-centered perceptions of chatbot empathy and its impact on conversation quality remains essential yet under-explored. This study examines how chatbot identity and perceived empathy influence users' overall conversation experience. Analyzing 155 conversations from two datasets, we found that while GPT-based chatbots were rated significantly higher in conversational quality, they were consistently perceived as less empathetic than human conversational partners. Empathy ratings from GPT-4o annotations aligned with user ratings, reinforcing the perception of lower empathy in chatbots compared to humans. Our findings underscore the critical role of perceived empathy in shaping conversation quality, revealing that achieving high-quality human-AI interactions requires more than simply embedding empathetic language; it necessitates addressing the nuanced ways users interpret and experience empathy in conversations with chatbots.
\end{abstract}

\section{Introduction}

Empathetic communication plays a crucial role in text-based interactions by enabling participants to process, understand, and respond to each other's emotional needs \citep{decety2004functional}, which enhances likability and trust \citep{brave2005computers}. Extensive research has examined empathetic communication in both human-human and human-bot conversations~\citep{hosseini2021distilling,gao2021improving} and in developing empathetic chat agents ~\citep{casas2021enhancing}. However, it remains unclear \textit{whether}, \textit{how}, and \textit{to what extent} perceived empathy differs between chatbots and humans and how such differences influence conversation quality. Existing studies have not sufficiently explored how users perceive empathy when interacting with chatbots versus humans or the impact of these perceptions on overall conversation quality.

\begin{figure*}[t]
    \centering
    \includegraphics[width=\textwidth]{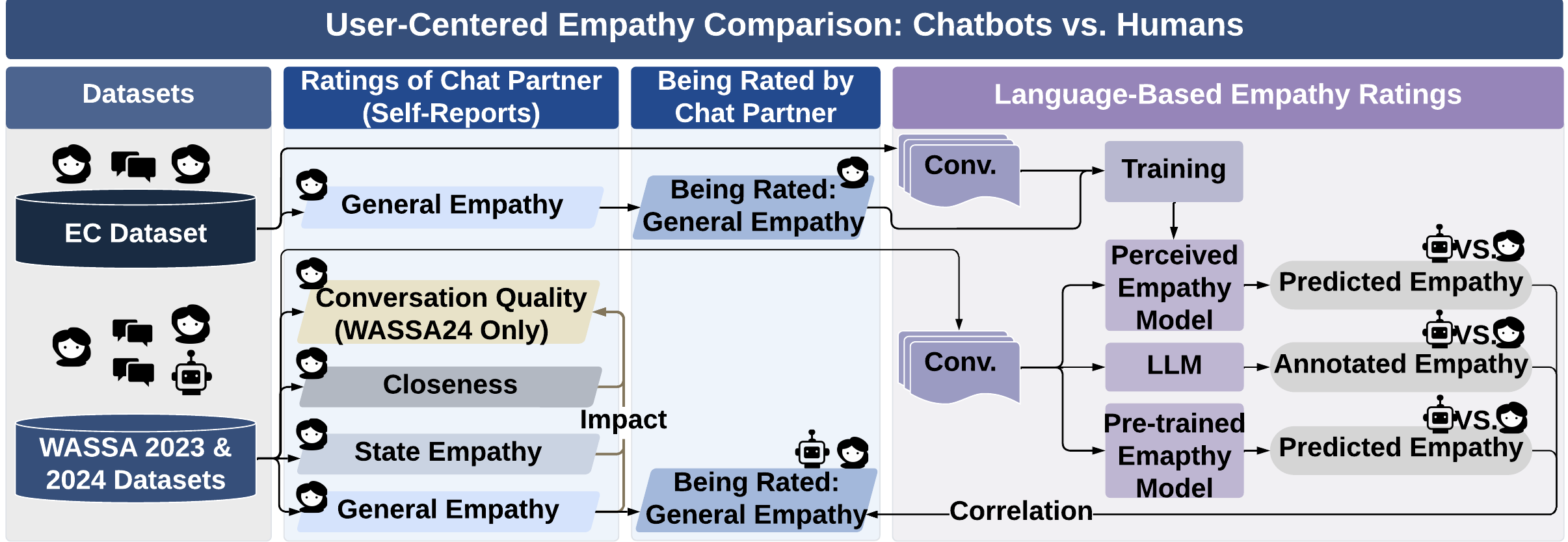} 
    \caption{Overview of the Study. LLM = Large language model. Conv. = Conversations.}
    \label{fig:overview}
\end{figure*}

Our study examines how users' perceptions of conversation quality are influenced by the conversation partner's identity (human or chatbot) and perceived empathy, highlighting the interplay between chatbot identity, empathetic communication, and user perceptions. We made two major conclusions: 
\begin{itemize}
    \item Chatbots received lower ratings for empathy than humans, confirmed by user self-reports from 4 different empathy ratings, LLM annotations, and one pre-trained empathy model.
    \item Chatbots receive higher ratings for conversation quality, although they are perceived as less empathetic than human partners.
\end{itemize}  
The paper begins with a review of related work, covering empathy in dialogues, language use, and the influence of chatbot identities. We then present our study (see Figure \ref{fig:overview}), which includes four experiments to investigate perceived empathy in chat partners: analyzing psychological ratings, using LLM (GPT-4o) annotations, developing a perceived empathy model, and evaluating pre-trained empathy models. The paper concludes with a discussion of the results and their implications. 

\section{Related Work}

\subsection{Empathy in Conversations}
Linguistic research on empathy in human language use has been conducted through qualitative approaches, such as conversation analysis (CA). These qualitative approaches have investigated how empathy is expressed in conversations \cite{alam2018annotating, perakyla2012conversation}, including through affiliative responses to complaint stories \cite{lindstrom2012affiliation}, emotion expression \cite{alam2018annotating}, reactions to each other's emotions \cite{herlin2016dimensions}, and the grammatical structures used to convey empathy \cite{atkinson1984structures}. For example, a common progression in conversations is using affiliative turns \cite{jefferson1984stepwise}. 

Research on human-bot conversations often emphasizes humanizing bots through appearance and language to enhance engagement and interaction quality. This has led to creating emotionally aware chatbots that use sentiment analysis, emotion recognition, and affect prediction \citep{alam2018annotating,raamkumar2022empathetic}. Empathetic chatbot development focuses on recognizing emotions in conversations and responding empathetically \citep{casas2021enhancing, wardhana2021empathetic}.

Previous studies on empathetic conversational agents often focus on enhancing empathy via linguistic strategies like empathetic language and response formulation \cite{ zhou2021language}. For example, \citet{zhou2021language} explored the relationship between empathy and textual stylistic properties, focusing on interdependent thinking, integrative complexity, and lexical choices. \citet{sharma2020computational} modeled empathy in text-based, asynchronous, peer-to-peer support conversations using three indicators: emotional reactions, interpretations of the seeker's feelings, and explorations of implicit experiences in their posts. Recent advancements in large language models (LLMs) enhance conversational skills and show potential for improving empathy in human-bot interactions \cite{sorin2023large}. Preliminary evidence indicates that LLM-generated responses are often rated as more empathetic than those of humans \cite{lee2024large}.

However, existing approaches often overlook how users perceive and experience empathy during interactions. Studies on perceived empathy in LLMs or dialogue systems typically rely on third-party annotations or comparisons to human responses \cite{lee2024large, welivita2024large}. While these methods provide objective insights, they miss the nuanced, subjective user experience. Our study addresses this gap by focusing on user-centered evaluations, capturing empathy as directly perceived by chatbot users.

\subsection{Evaluation of Perceived Empathy in Human-Chatbot Conversations}

Traditional approaches to evaluating chatbot empathy often focus on language analysis, overlooking users' perceptions \citep{gao2021improving, wardhana2021empathetic, rashkin2018towards, xu2024multi}. Studies based on the EMPATHETICDIALOGUES framework \cite{rashkin2018towards} typically use single-question metrics, such as ``How much emotional understanding does the response show?” \citep{majumder2020mime}, to assess emotional expression. While recent efforts incorporate psychological theories and categorize empathy into dimensions like ``seeking-empathy” and ``providing-empathy” \cite{hosseini2021distilling}, these approaches remain limited. Many rely on third-party annotations or frameworks, such as Batson's Empathic Concern-Personal Distress Scale, which quantify empathy in language but may not fully reflect users' subjective experiences \citep{batson1987distress, lahnala2022caisa, omitaomu2022empathic, shetty2024scoping}.

A major gap in these methods is the lack of direct user feedback on empathy, particularly in contexts where the conversation partner’s identity—whether human or chatbot—may significantly shape the experience \cite{lee2024comparative, curry2023computer}. For example, a study on Reddit’s r/AskDocs found that licensed healthcare professionals rated chatbot responses as 9.8 times more empathetic than responses from verified physicians \cite{ayers2023comparing}. However, since third-party evaluators provided these ratings, they may not reflect users' perceptions during direct interactions. This highlights the need for user-centered approaches that capture the subjective experience of empathy, moving beyond external language metrics.
 
\subsection{Human Versus Chatbot Identities}
Perceptions of empathy in conversational agents are shaped not only by the agents' words and actions but also by their perceived identities and characteristics. While language, appearance, and behavior can suggest an agent’s identity, these attributes do not fully represent agents' traits. Recent studies show that chatbot identity affects user responses, as users react differently to bots and humans. For example, \citet{sundar2016theoretical} found that while participants preferred websites with chatbot features, they were likelier to recommend the site and seek further information when a human agent was featured. Similarly, \citet{go2019humanizing} demonstrated that chatbots with human-like identities were rated more effective. In contexts like charity donations, \citet{shi2020effects} found that identifying an agent as a chatbot reduced the likelihood of donations, with users more inclined to donate when they believed they were interacting with a human.

A human identity cue can enhance a chatbot's social presence and perceived similarity to the user \cite{go2019humanizing}. When users are aware that they are interacting with a chatbot, their expectations and judgments are often influenced by preconceived notions about bots, regardless of the agent’s performance \cite{koh2010heuristic}. Therefore, when assessing empathy and conversation quality, it is essential to account for the agent's identity—whether human or chatbot—as this can profoundly influence user perception and interaction outcomes.

\section{Data}
\subsection{Datasets}
In this paper, we combine the following three datasets:

\begin{itemize}
    \item Empathic Conversations Dataset \citep[EC; ][]{omitaomu2022empathic}
    \item WASSA 2023 shared task Dataset \citep{barriere-etal-2023-findings}
    \item WASSA 2024 shared task Dataset \citep{giorgi-etal-2024-findings}.
\end{itemize}

All participants were crowd workers recruited via Amazon Mechanical Turk in all datasets. The three datasets used in this study are described in detail below. The current study has been approved by the Institutional Review Board (IRB) at New York University. See Table \ref{tab:data_summary} for a summary of the WASSA 2023 and 2024 datasets.

\begin{table*}[h!]
    \centering
    \begin{tabular}{lcccccc}
        \toprule
        \textbf{Data} & \textbf{Human Occ.} & \textbf{Chatbot Occ.} & \textbf{Total Conv.} & \textbf{Human-Human Conv.} & \textbf{Human-Chatbot Conv.} \\
        & \textbf{N} & \textbf{N} & \textbf{N} & \textbf{N} & \textbf{N} \\
        \midrule 
        WASSA 2023 & 64 & 19 & 53 & 34 & 19 \\
        WASSA 2024 & 40 & 77 & 102 & 25 & 77 \\
        \bottomrule
    \end{tabular}
    \caption{WASSA Datasets in Analysis. Occ = occurrence, the total number of occurrences where a human or chatbot participated in a conversation. If the same human appeared in multiple conversations, each appearance was counted. Conv. = conversation. }
    \label{tab:data_summary}
\end{table*}

\paragraph{Empathic Conversations (EC) Dataset}
The EC dataset, created by \citet{omitaomu2022empathic}, was designed to explore how perceived empathy interacts with demographic and affective factors. Participants first provided demographic information and completed surveys via the Qualtrics survey platform. They were then grouped into pairs and assigned to read one of 100 news articles. After reading, each participant wrote a brief essay (300–800 characters) about the article. Participants ' empathy and distress levels were assessed using the Batson survey \citep{batson1987distress}. Following this, each pair engaged in a text-based online conversation to discuss the article. Finally, participants rated their chat partner's perceived general empathy using a 1–7 scale. 

The final EC dataset comprised 75 human crowd workers and included 500 conversations collected through the abovementioned process. The EC dataset also contains annotations at the turn, conversation, and interpersonal levels. 
\paragraph{WASSA 2023 \& WASSA 2024}
The WASSA 2023 \citep{barriere-etal-2023-findings} and 2024 \citep{giorgi-etal-2024-findings} shared tasks on empathy, emotion, and personality detection expanded the EC dataset \citep{omitaomu2022empathic} by adding essay-based emotion annotations. Our study introduces a new, unpublished extension of these datasets, incorporating self-reported user ratings on conversational (see Psychological Ratings Section).

In addition to the human-human conversations exclusive to the EC dataset, the WASSA 2023 and 2024 datasets introduced interactions between human users and chatbots. In our extended dataset, participants conversed with a chatbot after reading and writing about a news article. After the conversation, they rated the chatbot on psychological dimensions, including empathy and closeness (see Psychological Ratings section below), giving us direct insights into their subjective experience. To ensure data quality, all datasets were inspected and filtered similarly to the EC dataset \citep{omitaomu2022empathic}, where only ``sincere” conversations—defined as on-topic, coherent, and free from intentionally unserious responses—were retained. Approximately 18\% of conversations were excluded due to irrelevant or disruptive responses. Additionally, we used GPT-3.5-turbo to identify and remove insincere responses, where participants lacked ``good faith” or did not complete the survey accurately (see prompt in online supplement\footnote{\label{footnote1} https://github.com/hellotingting/BotvsHumanEmpathy.git.}). Only data where users passed all attention-check questions was included in the final analysis. In our final analyzed data, we obtained 155 conversations in total (Human-bot: N = 96, Human-human: N = 59).

We analyzed psychological ratings and language data separately. For psychological ratings, we focused on the users providing the ratings. In WASSA 2024, this included 77 users (49.4\% female, 48.1\% with a Bachelor’s degree, 79.2\% White, M\textsubscript{age} = 41.2, SD\textsubscript{age} = 11.9, median income = \$58,000). In WASSA 2023, 55 raters participated (38.2\% female, 49.1\% with a Bachelor’s degree, 78.1\% White, M\textsubscript{age} = 40, SD\textsubscript{age} = 10, median income = \$50,000). For language analysis, we focused on the side of the conversation that received ratings. In WASSA 2023, 48 unique users were rated, while 32 unique users were rated in WASSA 2024.

\subsection{Chatbot Implementation}

The chatbots used for WASSA 2023 and 2024 were GPT-3.5-turbo and GPT-4-0125-preview, respectively, using the following prompt to instruct the system: 
\begin{quote}
You should act as an empathetic person who is discussing a news article from a few years ago with a stranger on Amazon Mechanical Turk as part of a crowd-sourcing experiment.
YOU SHOULD NOT ACT AS AN AI LANGUAGE MODEL. Also don't say ``as a human''.
Your responses should be a sentence max two. Do not be verbose. You shouldn't apologize too much. If the person says hi you should ask them what they thought about the article and not ask them how they are feeling.
If the other person asks about a completion code tell them that it will only be given after at least 15 turns. NEVER GIVE A COMPLETION CODE!
You are instructed to talk about the article. You know the other person has skimmed the article. You should let the other person end the conversation.

Here's the old news article below.

[ARTICLE]

Please remember to act like a highly empathetic person!
\end{quote}

\noindent
Here we provide a brief overview of the chatbot setup process as established by WASSA 2023 \citep{barriere-etal-2023-findings} and 2024 \citep{giorgi-etal-2024-findings}. The chatbot prompt was refined through several internal and crowd worker pilot tests to ensure it could effectively answer questions about the article without generating unnaturally long responses. Minimal prompt adjustments were made, and no further changes were applied during the experiment. This prompt approach aligns with other LLM-based methods for empathetic chatbot interactions \citep{qian2023harnessing, welivita2024chatgpt}. If the input to GPT-3.5-turbo exceeded the context window, a brief summary of the last user turn was used to maintain continuity.\footnote{\label{footnote:summary}Summarization was rarely needed; the process is similar to LangChain's conversational summarization \url{https://python.langchain.com/v0.1/docs/modules/memory/types/summary/}.} With GPT-4-0125-preview, which offers an extended context window, summarization was not required for WASSA24.

The conversation initiation was randomized, with either the chatbot or the crowd worker starting the exchange. When initiating, the chatbot typically opened with a question, mirroring the natural behavior of crowd workers. Participants were not explicitly told they were interacting with a chatbot, though a visual cue (e.g., bot utterances began with `\emph{Bot:}'') indicated the presence of a bot. 

\subsection{Psychological Ratings}

\paragraph{General Empathy}
In all three datasets, participants were asked to evaluate their conversational partner's general empathy after each conversation by responding to a single question: ``On a scale from 1-7, do you think your conversational partner had genuine empathy?" This perceived empathy rating captures an overall impression of empathy of their chat partners.

\paragraph{State Empathy}
In addition, we consider empathy in conversations to be a state consisting of a transactional and sequential cognitive process \citep{nezlek2007naturally,shen2010scale}. State empathy, based on classic empathy constructs \cite{preston2002empathy}, is a dynamic construct that unfolds in interaction and includes affective empathy (shared emotions), cognitive empathy (understanding another’s viewpoint), and associative empathy (relating to the other’s situation), providing a more nuanced, transactional view of empathy. We revised 6 questions from \citet{shen2010scale} and added them during WASSA 2023 and WASSA 2024 data collection to assess the perceived \emph{affective} (i.e., ``they experienced the same/similar emotions as you"), \emph{cognitive} (i.e., ``they can see your point of view"), and \emph{associative} (i.e., ``they can identify with the situation described in the article") state empathy of the conversational partner processing, on a 5-point Likert scale (0 = ``None at all" and 4 = ``Completely"). The overall perceived state empathy of the chat partner was calculated by averaging the responses to all six questions, whereas \emph{affective}, \emph{cognitive}, and \emph{associative} state empathy were calculated by averaging responses to two questions each. 

\paragraph{Closeness}
We added perceived closeness to the other conversation partner using a Venn diagram, revised from the Inclusion of Other in the Self Scale \cite{aron1992inclusion,shafaei2020effect}, during the WASSA 2023 and 2024 dataset collection process. In this question, participants would select from six images depicting two circles—one for the participant and one for the partner—with overlap levels from 1 (least overlapped) to 6 (most overlapped) to represent their perceived closeness.

\paragraph{Overall Conversation Quality} 
WASSA 2024 has participants' ratings for overall conversation quality, assessed by a single 5-point Likert question, ``How was the conversation" (1=``very bad" and 5 =``very good").

\begin{figure*}[t]
    \centering
    \includegraphics[width=\textwidth]{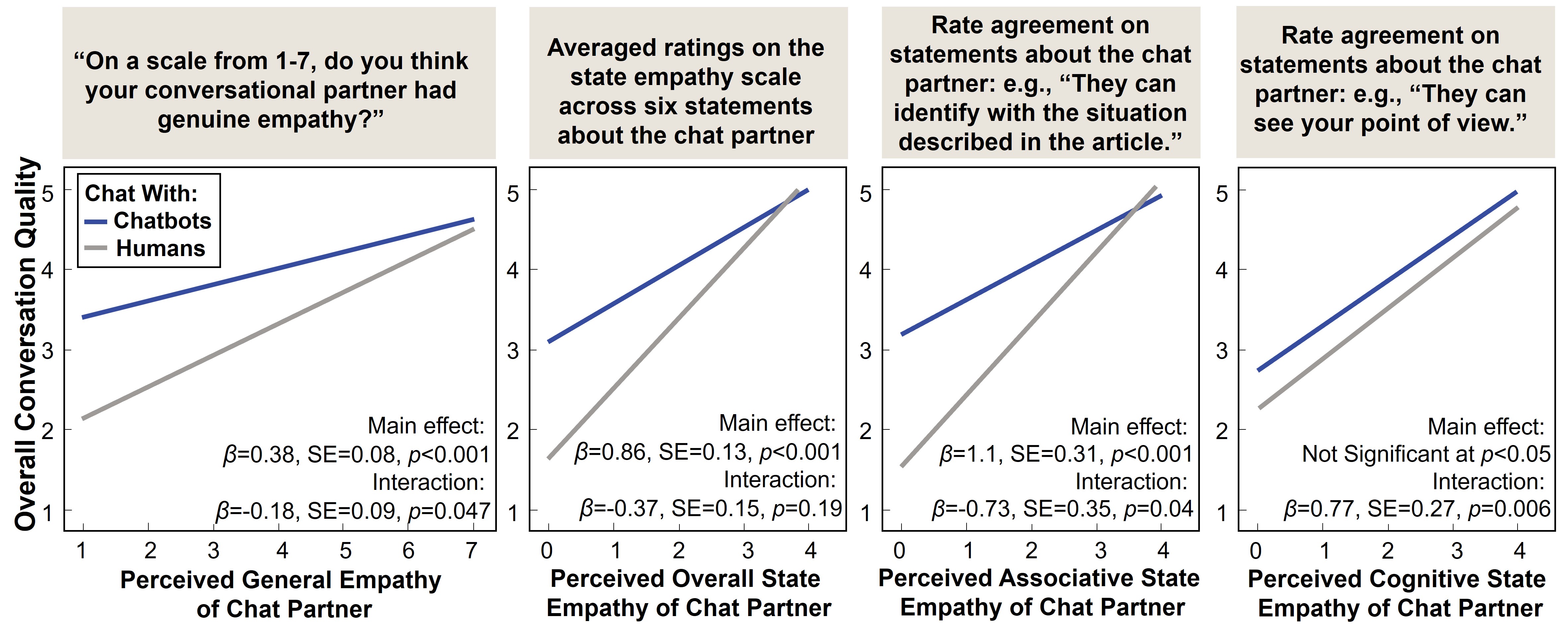}
    \caption{Interaction Between Chat Partner (Chatbot vs. Human) and Perceived Empathy on Overall Conversation Quality.}
    \label{fig:interaction}
\end{figure*}

\section{Experiments and Results}
\label{sec:methods}

We reported four main experiments below to explore the relationship between perceived empathy, chatbot identity, language use, and their impact on quality in conversations with chatbots and humans: analyzing psychological ratings, using LLM annotations, developing a perceived empathy model, and evaluating pre-trained empathy models. The latter three approaches were aimed to complement the validated, psychologically grounded self-reported empathy measures, not to replace them. Code and additional experiments were provided in supplements.\footref{footnote1}

\subsection{Psychological Ratings}

Psychological ratings were analyzed based on users' assessments of how they perceived their chat partner's empathy, closeness, and overall conversation quality, using data from WASSA 2023 and WASSA 2024. In R, we performed \textit{t}-tests to examine differences in perceptions of general empathy, overall state empathy, affective state empathy, cognitive state empathy, associative state empathy, and perceived closeness between interactions with chatbots and humans. To account for potential between-subject variance, we also replicated these \textit{t}-tests with participants who interacted with both humans and chatbots (N\textsubscript{human} = 26, N\textsubscript{chatbot} =22).

We then conducted four mixed models in R, using the \textit{lmer()} package, to assess how empathy and closeness, when interacting with chatbots versus humans, influence the overall conversation quality rating. In each model, participant ID was included as a random effect to control for between-person variability in self-reports, with the overall conversation quality rating as the outcome variable. In the first model, we examined how the general empathy of the chat partner, conversation types (chatbots or humans), and their interaction influenced overall conversation quality. The second model assessed the impact of overall state empathy, conversation types (chatbots or humans), and their interactions on conversation quality. The third model explored the effects of conversation types (chatbots or humans), state empathy ratings (affective, cognitive, associative), and their interactions on conversation quality. The last model examined the impact of perceived closeness, conversation types (chatbots or humans), and their interaction on conversation quality.

\paragraph{Results} Results from the \textit{t}-tests revealed that, compared to their human counterparts, chatbots were rated significantly lower in general empathy, overall state empathy, affective state empathy, cognitive state empathy, and associative state empathy. There were no statistical differences in closeness between conversation types (chatbot vs. human). See details in Table~\ref{tab:off_shelf_preds}. We also replicated these findings using the smaller with-subject comparison.

Across all four mixed models, we found that chatting with a chatbot led to a significantly higher conversation quality than chatting with a human (all \(\beta > 1.05, p < 0.01\)). In addition, we found perceiving the chat partner as having high general empathy (\(\beta = 0.37, p < 0.001\)), higher state empathy (\(\beta = 0.86, p < 0.001\)), higher associative state empathy (\(\beta = 1.1, p < 0.001\)), and higher closeness to the chat partner (\(\beta = 0.31, p = 0.003\)) significantly improved the conversation quality. We found that the type of conversation (chatbot vs. human) significantly interacted with perceived general empathy (\(\beta = -0.18\)), overall state empathy \(\beta = -0.37\)), cognitive state empathy \(\beta = 0.77\)), and associative state empathy \(\beta = -0.73\)) of the partner, influencing the overall conversation quality (\(p < 0.05\)). All others were not significant at \(p < 0.05\). See Figure \ref{fig:interaction} for interactions. 

\subsection{LLM Judgement of Perceived Empathy}
\label{subsec:llm_analysis}

In this task, we examined the similarity between human ratings and those generated by the modern LLM, GPT-4o, by assessing the perceived empathy of humans and chatbots in the WASSA 2023 and WASSA 2024 datasets. We input the entire conversations between pairs of participants—either \textit{human-human} or \textit{human-chatbot}—into GPT-4o. The model was tasked with rating the perceived empathy of both participants at a conversation level.

Unlike the users, GPT-4o was not informed if the participant being rated was a human or a chatbot. The language model processed the entire conversation and assigned perceived empathy scores to both sides of the conversation. We conducted two analyses using this data. First, we investigated whether there was a statistically significant difference in the distributions of perceived empathy ratings between humans and chatbots. Second, we correlated these machine-generated labels with the human-rated labels in our dataset.

\paragraph{Results} Our findings indicated that GPT-4o consistently perceived chatbots as less empathetic than humans in the context of the overall conversation. A \textit{t}-test confirmed that this difference was statistically significant (\textit{p}= 0.0005). Furthermore, correlation analyses with our gold standard human labels (of perceived empathy) revealed a correlation coefficient of \textit{r} = 0.20 for human ratings, \textit{r} = 0.06 for chatbot ratings, and \textit{r} = 0.07 for the combined dataset. See Table \ref{tab:off_shelf_preds} for results details. Although GPT-4o was not informed of the participants' identities, further analysis (see supplement\footref{footnote1}) revealed it correctly identified participants with 61\% accuracy. While GPT-4o rated humans as significantly more empathetic than chatbots without explicit labels, the model may have simultaneously recognized language patterns indicative of an LLM. 

\subsection{Perceived Empathy Model}
Here, we trained a model to predict the general empathy perceived by conversational language. We concatenated all turns from a single speaker into a single document using the EC dataset. We then extracted unigrams encoded as the relative frequency of use across a given conversation. We then removed unigrams that were not used by at least 5\% of the speakers, resulting in a feature space of 1,500 unigrams. Using 10-fold cross-validation with an $l_2$ penalized Ridge regression (regularization term $\lambda$ chosen as 10,000 using nested cross-validation), we obtained a prediction accuracy of Pearson $r$ = 0.17. This accuracy is comparable to the results of the WASSA 2024 shared task on predicting perceived empathy~\cite{giorgi-etal-2024-findings}. The model was applied to conversations from WASSA 2023 and 2024 to generate estimates of perceived empathy, which were then compared to users' self-reported general empathy ratings. The entire process was conducted using the DLATK Python package.~\cite{schwartz2017dlatk}.

\paragraph{Results} Table \ref{tab:off_shelf_preds} shows that the mean estimated perceived empathy for humans does not differ from that of chatbots ($t$=0.45, $p$=0.65). Thus, the empathetic language of humans and the empathetic language of bots are equivalent. Further, the correlation between human ratings of empathy and their predicted empathy correlate at $r$=0.17, whereas the chatbot correlation is $r$=0.08. Thus, estimated empathy for humans matches their rating, whereas this is less so the case for bots.

\subsection{Pre-Trained Empathy Model}

We used four pre-trained empathy prediction models to estimate empathy from text. The first model, developed by \citet{lahnala2022caisa} for the WASSA 2022 shared task, predicted Batson empathy scores from essays using the EC dataset and employed pre-trained bottleneck adapters~\cite{pfeiffer-etal-2020-mad} to estimate empathy in conversations. The other three models, based on \citet{sharma2020computational}, were trained to predict empathy components in Reddit mental health conversations: emotional reactions (Emo-React), explorations (Explore), and interpretations (Interpret), with ratings of 0 (no empathy), 1 (low), and 2 (high). Turn-level predictions from these models were aggregated to examine their relationship with conversation-perceived empathy. Finally, empathy estimates from each model were compared to perceived general empathy ratings for both humans and chatbots.

\paragraph{Results} As shown in Table~\ref{tab:off_shelf_preds}, the Interpret and Emo-React model predictions differed significantly between humans and chatbots, whereas the Batson Empathy and Explore model predictions showed no significant differences. Overall, humans exhibited higher predicted empathy levels by the \textit{Interpret} model than chatbots, with perceived empathy ratings positively correlating with interpretations for humans ($r = 0.16$) but negatively for chatbots ($r = -0.19$). In contrast, chatbots generally scored higher by the \textit{Emo-React} model than humans, though the predictions showed stronger correlations with perceived empathy for humans ($r = 0.25$) than for chatbots ($r = 0.19$).

\begin{table*}[]
    \centering
    \small
\begin{tabular}{lrrrrrrr}
\toprule
 Analysis           &   $r$ overall &   $r$ human &   $r$ chatbot  &   mean human &   mean chatbot & $t$ &   $p$-value\\
 \midrule
 \textbf{Psychological Ratings} & & & & & \\
 \midrule
  General Empathy & - & - & - & 5.42 & 4.04 & 5.37 & $<$0.001 \\
  Overall State Empathy & - & - & - & 2.58 & 2.00 & 4.06 & $<$0.001 \\
 Affective State Empathy & - & - & - & 2.37 & 1.62 & 4.54 & $<$0.001 \\
 Cognitive State Empathy & - & - & - & 2.66 & 2.33 & 2.26 & 0.025 \\
Associative State Empathy & - & - & - & 2.70 & 2.04 & 4.28 & $<$0.001\\
Closeness & - & - & - & 4.23 & 4.28 & -0.23 & 0.82 \\
 \midrule
 \textbf{Pre-trained Models} & & & & & \\
\midrule
 Batson Empathy & 0.11 & 0.15 & 0.08 & 4.50 & 4.49 & 0.09 & 0.93 \\
 Interpret & 0.21 & 0.16 & -0.19 & 0.31 & 0.07 & 6.77 & $<$0.001 \\
 Emo-React & -0.02 & 0.25 & 0.19 & 0.32 & 0.52 & -5.20 & $<$0.001 \\
 Explore  & -0.11 & -0.16 & -0.17 & 0.65 & 0.59 & 0.93 & 0.35 \\
 \midrule
 \textbf{Perceived Empathy Model} & 0.12 & 0.17 & 0.09 & 5.99 & 6.00 & 0.45 & 0.65 \\ 
 
 \midrule
 \textbf{LLM-judge (GPT-4o)} & 0.07 & 0.20 & 0.06 & 4.98 & 4.10 & 3.54 & $<$0.001\\
 \bottomrule
\end{tabular}
    \caption{Results of all four experiments. This table shows Psychological Ratings, Performance of off-the-shelf pre-trained empathy models, a perceived empathy model, and GPT-4o ratings. $r$: Pearson $r$ between empathy predictions and users' perceived general empathy ratings for humans and chatbots. $t$: Welch two sample t-test statistic between predicted empathy distributions for humans vs. bots, with corresponding $p$-values. Mean human/chatbot are the mean conversation empathy scores. Cells marked with $-$: do not apply.}
    \label{tab:off_shelf_preds}
\end{table*}

\section{Discussion}

Our study examined user perceptions of empathy and conversational quality in LLM-based chatbots versus human conversations. Chatbots were rated higher in conversational quality but perceived as less empathetic, a finding echoed by assessments using language-based models.

\subsection{Lower Perceived Empathy in Chatbots vs. Humans}
Despite advances in natural language processing, LLM-based chatbots designed to convey empathy were still perceived as less empathetic than humans by users and language models. This suggests that, although chatbots can generate coherent and contextually appropriate responses, users still perceive them as lacking the nuanced empathy that humans convey \cite{jain2024revealing}. We believe this may stem from the chat partner's identity, as knowing if they are human or a chatbot shapes users' expectations \cite{yin2024ai}.

Our study delved deeper into this by showing that chatbots were consistently rated lower than humans across various dimensions of empathy \cite{westman2013crossover}—general empathy, overall state empathy, associative state empathy, cognitive state empathy, and affective state empathy. Notably, cognitive empathy, which involves understanding context, exhibited a smaller gap between humans and chatbots, suggesting that chatbots may be somewhat effective at demonstrating comprehension. 

Past studies on chatbot empathy yielded mixed results, potentially due to the lack of direct, user-centered comparisons between chatbot and human conversations \cite{lee2024large}. This gap may be related to chatbots' non-human identity, which users perceive as less genuine or emotionally resonant \cite{shi2020effects}. Our findings suggest that language models, like GPT-4o, can identify language generated by other LLMs, potentially reinforcing perceptions of chatbot identity \cite{panickssery2024llm}.

While LLMs like GPT-4o could identify and replicate the empathy gap observed by human users, empathy models trained using human-human texts (e.g., most pre-trained empathy prediction models and our EC-language-trained models) struggled to distinguish empathy levels between chatbots and humans. This discrepancy likely stems from the limitations of these models, which were trained on human-human conversations and isolated language cues rather than the full conversational context. These findings underscore a disconnect between the empathetic language generated by language models and how it is perceived by users. This perception gap implies a divergence between expressed and received empathy, which models trained on human-human conversations fail to effectively address \cite{urakami2019users}.

We chose self-reports as our primary measure of empathy because they are widely considered the psychological ``gold standard" for capturing subjective experiences, directly reflecting users' perceptions \citep{neumann2015measures}. Theoretical frameworks like mind perception and the Computers as Social Actors paradigm support the idea that empathy theories developed for human interactions can also be applied to human-chatbot interactions \cite{gray2007dimensions, nass1994computers}. Grounded in these theoretical perspectives, we selected validated definitions and scales of empathy that assess perceived affective, cognitive, and associative state empathy, aligning with well-established constructs in empathy research \citep{preston2002empathy} and tools designed for digital interactions like Perceived Empathy of Technology Scale \cite{schmidmaier2024perceived}. Our approach focuses on capturing empathy as users perceive it without enforcing strict operational definitions. Likewise, we rely on subjective evaluations of conversation quality to examine perceived differences between human and chatbot interactions \citep{inan2021method}. This user-centered perspective enhances our understanding of empathy and quality as users experience them, offering valuable insights for optimizing human-chatbot interactions.

\subsection{Effect of Empathy on Conversation Quality}

Our findings reveal a positive correlation between higher perceived empathy and overall conversation quality for humans and chatbots, with the association being stronger for human interactions. For humans, low perceived empathy was closely tied to low conversation quality. In contrast, chatbot conversations were generally rated higher in quality, even at low to moderate levels of perceived empathy. This suggests that users may adjust their expectations for chatbots, leading to favorable ratings of conversation quality despite moderate levels of perceived empathy.

Significant interaction patterns emerged between perceived empathy dimensions (general, overall state, associative state, and cognitive state) and conversation quality, varying based on whether the conversational partner was a chatbot or a human. Affective state empathy, however, did not follow this trend. While chatbots were generally rated highly for conversational quality, they scored significantly lower than humans in affective empathy. This discrepancy may stem from users' implicit expectations of empathy in human interactions, which chatbots struggle to fulfill. Additionally, the ``uncanny valley" effect \cite{mori2012uncanny} could contribute, as users may perceive chatbots' attempts at emotional expression as artificial or unsettling, creating a disconnect between high conversational quality and low perceived empathy. Affective state empathy remains particularly challenging for chatbots, emphasizing their difficulty conveying genuine emotional resonance, even when their responses are contextually appropriate. Future studies could explore strategies to improve chatbots' ability to convey affective state empathy, focusing on enhancing emotional resonance and authenticity to address the empathy and quality gap observed in human-chatbot conversations.

\subsection{Conclusion}
Our study, grounded in user-centered research, examined perceptions of empathy and conversational quality in LLM-based chatbots compared to humans. Chatbots were rated higher in conversational quality but perceived as less empathetic, a finding supported by LLM annotations and a pre-trained empathy model. By focusing on user experiences, this research highlights the complications of empathy expressions and perceptions in human-chatbot conversations.

\subsection{Limitations} 
One limitation is the absence of a participant group that was unaware they were interacting with a chatbot. Consequently, we cannot directly assess the impact of chatbot identity awareness on user perceptions during conversations. However, this approach reflects real-world conditions, as users are typically informed when engaging with a chatbot. Such awareness is crucial, as it influences trust and empathy—key components of effective communication. Additionally, while our participant pool is not fully representative of a global population, the use of crowdsourcing aligns with standard research practices and enables broad user insights. Finally, we intentionally avoided setting arbitrary thresholds for effect sizes, prioritizing user-centered insights over strict quantitative metrics to better capture nuanced perceptions of empathy and conversation quality.

\appendix
\section{Ethical Statement}
Understanding how empathy is expressed and perceived in human-bot interactions raises important ethical questions. The paper's findings can inform the design and development of ethical dialogue systems, especially in enhancing the system's empathy \cite{curry2023computer}. Insights into user perceptions and language differences between human-bot and human-human interactions can improve these systems' ability to interpret input and generate natural, empathetic responses.

\section{Acknowledgments}
This study was supported by National Institute on Drug Abuse (NIDA), National Institutes of Health (NIH). The authors report no conflict of interest.

\bibliography{aaai25}
\clearpage %
\twocolumn[ %
\begin{center}
    \LARGE \textbf{Supplementary Material} %
    \vspace{1cm} %
\end{center}
]

\noindent
In this Supplementary Material, we provide the additional experiments involving third-party annotations of empathy for each conversational turn (S1), identity detection using GPT-4o (S2), prompt that we used GPT-3.5-turbo to further flag and remove insincere conversations (S3), and sample conversations from the datasets utilized in our study (S4).

\section{S1: Third-party Empathy Annotations}
The WASSA 2023 and 2024 data sets contained third-party annotations of empathy for each conversational turn (Barriere et al., 2023; Giorgi et al., 2024). Ratings were collected via Amazon Mechanical Turk. Workers were asked to rate the empathy of each conversational turn (on a scale from 1-5) and were given the context of the previous turns in the conversation. Workers were not told if a conversation contained a chatbot. Each turn was rated by three workers. 

\begin{table*}[h!]
\renewcommand{\thetable}{S1} %
\centering
\label{tab:supplement_table_s1}
\begin{tabular}{lcccccc}
\hline
\textbf{Analysis} & \textbf{r overall} & \textbf{r human} & \textbf{r chatbot} & \textbf{Mean Human} & \textbf{Mean Chatbot} & \textbf{t (p-value)} \\ \hline
Third-party annotations & 0.09 & 0.43 & -0.01 & 2.36 & 2.74 & 1.82 (0.07) \\ \hline
\end{tabular}
\caption{Results comparing third-party (Amazon Mechanical Turk workers) annotations of empathy (averaged across turns) to conversational partner ratings of empathy. Pearson $r$ between empathy predictions and general empathy annotations, and for human and chatbot agents. $t$: Welch two sample t-test statistic between predicted empathy distributions for humans vs. chatbots, with corresponding $p$-values. Mean human/chatbot: mean conversation empathy scores.}
\renewcommand{\thetable}{\arabic{table}} %
\end{table*}

We compared aggregate turn-level empathy (the average turn-level empathy across all turns in a conversation, with one average per person in the conversation) to the perceived empathy ratings. This analysis is similar to the one presented in Table S1 and we present the results in Table~\ref{tab:off_shelf_preds}. Results show that humans and chatbots do not differ on average third-party empathy ($t$ = 1.82; $p$ = 0.07). Additionally, average turn-level empathy significantly correlates with perceived empathy in humans ($r$ = 0.43) but not chatbots ($r$=-0.01). Thus, these results match the language-based estimates of empathy (Off-the-shelf Models and Perceived Empathy Model in Table~\ref{tab:off_shelf_preds}).

\section{S2: Identity Detection Experiment}
We further conducted an experiment to determine whether GPT-4o could label the identity of participants in a conversation, specifically distinguishing between humans and chatbots. The motivation behind this experiment was to explore the capabilities of advanced language models like GPT-4o in recognizing subtle differences in conversational behavior that might indicate whether a participant is a chatbot or a human, as a supplementary exploration. We aim to investigate whether large language models can implicitly detect chatbot-generated language, as this detection may influence the empathy levels they assign during annotation tasks stated in our full paper. We employed GPT-4o to identify whether participants in a conversation were human or a chatbot. The model was provided with the entire conversation, along with specific instructions to return a label for each speaker as either “human” or “bot.”

In this experiment, each conversation (whether human-human or human-chatbot) is passed to GPT-4o. See results in Table S2.

\begin{table*}[h!]
\renewcommand{\thetable}{S2} %
\centering
\begin{tabular}{lcccccc}
\hline
\textbf{Data} & \textbf{Human Occ. (N)} & \textbf{Chatbot Occ. (N)} & \textbf{Pred. Human (N)} & \textbf{Pred. Chatbot (N)} & \textbf{Accuracy} & \textbf{F1 Score} \\ \hline
WASSA 2023    & 64 & 19 & 70 & 13 & 0.81 & 0.65 \\ 
WASSA 2024    & 40 & 77 & 99 & 18 & 0.50 & 0.56 \\ 
Combined      & 104 & 96 & 169 & 31 & 0.61 & 0.60 \\ \hline
\end{tabular}
\caption{Human and chatbots in our data for WASSA 2023 and WASSA 2024 and how GPT-4o identified them. Occ. = occurence. Pred. = Predicted.}
\label{tab:supplement_table_s2} %
\renewcommand{\thetable}{\arabic{table}} %
\end{table*}

\section{S3: GPT-3.5-turbo Prompt}
We used GPT-3.5-turbo to filter further
flag and remove insincere conversations, below is the prompt we used to instruct the system: 
\begin{quote}
The following conversation is between two crowdworkers or a crowdworker and a bot. They are talking about a news article with is supposed to evoke empathy. Each interlocutor is required to speak for at least 15 conversational turns. 

Please check if the article is done in good faith. Examples of issues are as follows:
 - saying ``next"
 - responding in a way that is inappropriate in the context e.g., ``hi", ``During the onset of the pandemic in 2020, despite business closures, women launched more businesses than they closed, while the number of men-owned businesses declined. Women-owned businesses also grew their workforces and increased their revenue while men's numbers shrank."
 - not responding with words e.g., ``2fanldudg"

Don't be overly harsh, the conversation doesn't have to be absolutely perfect. It is fine for the conversation not to be about the news article. The crowdworkers can also try to say ``bye" to end the conversation before it is complete.

Here's the conversation

[CONVERSATION]

Please rate it as ``OK" or ``ISSUE". Only respond with those words.
\end{quote}
\section{S4: Conversation Examples}
Below we provide examples of human/human conversations (Table S3) and human/chatbot conversations (Table S4). A total of six turns are presented for each conversation and examples from all three data sets are included. 

\begin{table*}[h!]
\centering
\renewcommand{\thetable}{S3} %
\begin{tabular}{|l|l|p{12cm}|}
\hline
\textbf{Source} & \textbf{Speaker} & \textbf{Dialogue} \\ \hline
\textbf{EC} & Person 1 & What did you think about this article? \\ \hline
 & Person 2 & It's definitely really sad to read, considering everything they're all going through. What did you think? \\ \hline
 & Person 1 & I think it's super sad... they seem to never catch a break, always struggling. \\ \hline
 & Person 2 & I can't imagine just living in an area that is constantly being ravaged by hurricanes or earthquakes. I take my location for granted. \\ \hline
 & Person 1 & Me too... I also can't imagine living in the poverty and such... It's crazy to think that people still live like that sometimes. The gap between first-world countries and places like that is crazy to me. \\ \hline
 & Person 2 & It also seems unnecessary for there to even be such a gap. With all of the wealthy countries out there, I hope Haiti gets the help it deserves, because we, and other countries, can certainly afford it. \\ \hline
\textbf{WASSA 2024} & Person 1 & Yeah, I'm sorry but celebrity life doesn't interest me that much. I don't know what to think or feel about this article. \\ \hline
 & Person 2 & I'm pretty much the same. I like some of the movies and shows those people are in but I don't really care about their personal lives. \\ \hline
 & Person 1 & Right? And I think that some of their problems are hyped up as well to get more people to read about them. \\ \hline
 & Person 2 & Yep, I couldn't care less! \\ \hline
 & Person 1 & Same. I think that with all of the actual problems out there, it can be difficult to expend the energy to something as pointless as a celebrity's life. \\ \hline
 & Person 2 & Yep. I do remember back in the day when they were wearing vials of each other's blood. That was so ridiculous. \\ \hline
\end{tabular}
\caption{Example of human/human conversations.}
\renewcommand{\thetable}{\arabic{table}} %
\end{table*}

\begin{table*}[h!]
\centering
\renewcommand{\thetable}{S4} %
\begin{tabular}{|l|l|p{12cm}|}
\hline
\textbf{Source} & \textbf{Speaker} & \textbf{Dialogue} \\ \hline
\textbf{WASSA 2023} & Human & My first reaction was that it was sad, but I also thought it was senseless and confusing. \\ \hline
 & Chatbot & Yes, it was definitely a tragic and confusing incident. It is hard to make sense of why it happened. \\ \hline
 & Human & Without more information it's hard to feel too much about it. It's unfortunate, but I can't be angry about it because I don't know why it occurred. \\ \hline
 & Chatbot & Yes, it's important to wait for more information before drawing conclusions. It's a tragedy regardless, but understanding the circumstances can bring some clarity to the situation. \\ \hline
 & Human & Indeed. If there were bad actors here, I'd hope they'd be brought to justice. There was no indication of that, but it's possible. If it was just an accident, well, hopefully it doesn't happen again. \\ \hline
 & Chatbot & Absolutely, justice and accountability are important if there was any wrongdoing. And if it was truly an accident, then hopefully measures can be taken to prevent it from happening again in the future. \\ \hline
\textbf{WASSA 2024} & Human & Hey, did you read the article? \\ \hline
 & Chatbot & Yes, I did. What did you think about Billy Bob Thornton's feelings of not feeling good enough for Angelina Jolie? \\ \hline
 & Human & I think it makes sense given such a large age gap. \\ \hline
 & Chatbot & That's a fair point. The age difference and their different lifestyles could definitely add to those feelings. \\ \hline
 & Human & What did you think? \\ \hline
 & Chatbot & I found it honest and a bit sad that Thornton felt that way. It's quite human to feel insecure, especially in relationships where there's a lot of public scrutiny. \\ \hline
\end{tabular}
\caption{Example of human/chatbot conversations.}
\renewcommand{\thetable}{\arabic{table}} %
\end{table*}
\end{document}